\documentstyle[aps,epsfig]{revtex}
\begin{document}
\draft
{
\title{Energy dependence of the quark masses and mixings\thanks{
Talk presented at the \textit{IX Mexican School on Particles and
Fields}, August 9--19, Metepec, Pue., Mexico. To be published in the
AIP Conference Proceedings.}}
\author{S.R.~Ju\'{a}rez~W.$^{\dagger}$, S.F.~Herrera H.$^{\dagger}$, 
P.~Kielanowski$^{\ddagger}$
and G.~Mora H.$^{\ddagger }$}
\address{$^{\dagger}$Departamento de F\'{\i}sica,
Escuela Superior de F\'{\i}sica y
Matem\'{a}ticas, IPN, M\'{e}xico\\
$^{\ddagger }$Departamento de F\'{\i}sica,
Centro de Investigaci\'{o}n y Estudios Avanzados, M\'{e}xico\\
$^{\ddagger }$Institute of Theoretical Physics, University of
Bia{\l}ystok, Poland}
\maketitle
\begin{abstract}
The one loop Renormalization Group Equations for the Yukawa couplings
of quarks are solved. From the solution we find the explicit energy
dependence on $t=\ln E/\mu $ of the evolution of the {\em down} quark
masses $q=d,s,b$ from the grand unification scale down to the top
quark mass $m_{t}$. These results together with the earlier published
evolution of the {\em up}\/ quark masses completes the pattern of the
evolution of the quark masses. We also find the energy dependence of
the absolute values of the Cabibbo-Kobayashi-Maskawa (CKM) matrix
$|V_{ij}|$. The interesting property of the evolution of the CKM
matrix and the ratios of the quark masses: $m_{u,c}/m_{t}$ and
$m_{d,s}/m_{b}$ is that they all depend on $t$ through only one
function of energy $h(t)$.
\end{abstract}
}
In a recent paper Ref.~\cite{ref0}, a systematical investigation of
the evolution of the CKM matrix and the quark Yukawa couplings
\thinspace $y_{u}(t)$ and $y_{d}(t)$ was performed. Exact solutions of
the one loop Renormalization Group Equation (RGE) and some general
properties of the RGE evolution for the quark masses $m_{q}$,
$q=u,c,t,d,s,b,$ and CKM matrix, compatible with the observed
hierarchy Ref.~\cite{W}, were obtained.

The objective of this talk is to present some complementary results to
the previous ones. We show here the explicit solutions for the masses
of the quarks of the down sector and how they are derived.

The quark masses are the eigenvalues of the Yukawa couplings obtained
after its diagonalization by biunitary transformations with the help
of the unitary matrices $(U_{u,d})_{L,R}$
\[
\mbox{Diag}(m_{u},m_{c},m_{t})=(U_{u})_{L}y_{u}(U_{u})_{R}^{\dagger },
\,\,
\mbox{Diag}(m_{d},m_{s},m_{b})=(U_{d})_{L}y_{d}(U_{d})_{R}^{\dagger }.
\]
From the diagonalizing matrices we obtain the flavor mixing in the
charged current described by the CKM matrix
\[
V_{\mbox{\scriptsize CKM}}=(U_{u})_{L}(U_{d})_{L}^{\dagger }.
\]
The Yukawa couplings are scale dependent. Very frequently one makes
various assumptions about their properties at the Grand Unification
(GU) scale and then one has to compare the predictions with the
measured values at low energies Ref.~\cite{ref1}. The RGE are an
important tool for the search of the properties of the quark masses
and the CKM matrix at different energy scales. The RGE have been
worked out by various authors Ref.~\cite{ref2,ref3,ref4,ref7}. The
structure of the one loop RGE for the gauge coupling constants $g_{k}$
and the Yukawa couplings $y_{u,d,e,\nu }$ is the following
\[
\frac{dg_{k}}{dt}=\frac{1}{(4\pi)^{2}}b_{k}g_{k}^{3},
\,\,\,\,\,\,\,\,\,\,
\frac{dy_{u,d,e,\nu }}{dt}=\left[\frac{1}{(4\pi)^{2}}
\beta _{u,d,e,\nu}^{(1)}\right] y_{u,d,e,\nu }.
\]
Here $t\equiv \ln (E/\mu)$ is the energy scale parameter, the
coefficients $b_k$ are defined in Table~1 and the functions
$\beta_{u,d,e,\nu}^{(1)}$ are defined for various models in the
Appendix. The approximate form of the equations for the quark Yukawa
couplings, neglecting all the terms of $\lambda^{4}$ and higher
($\lambda =0.22$) have the following form
\begin{eqnarray}
\frac{dy_{u}}{dt} &=&\frac{1}{(4\pi)^{2}}[\alpha _{1}^{u}(t)
+\alpha_{2}^{u}y_{u}^{\phantom {\dagger }}y_{u}^{\dagger }
+\alpha _{3}^{u}\mbox {Tr}(y_{u}^{\phantom {\dagger }}
y_{u}^{\dagger })]y_{u},  \nonumber \\
\frac{dy_{d}}{dt} &=&\frac{1}{(4\pi)^{2}}[\alpha _{1}^{d}(t)
+\alpha_{2}^{d}y_{u}^{\phantom {\dagger }}y_{u}^{\dagger }
+\alpha _{3}^{d}\mbox {Tr}(y_{u}^{\phantom {\dagger }}
y_{u}^{\dagger })]y_{d}.
\end{eqnarray}
The explicit solutions for $m_{q}(t)$ with $q=u,c,t$ and $y_{d}(t)$
previously obtained in Ref.~\cite{ref0}\thinspace are:
\begin{eqnarray}
&m_{u,c}(t)=m_{u,c}(t_{0})\,\sqrt{r_{g}(t)}(h(t))^{b/c},
\,\,\,\,\,\,\,\,\,\,
m_{t}(t)=m_{t}(t_{0})\sqrt{r_{g}(t)}\,(h(t))^{\frac{(b+2)}{c}},
\nonumber \\
&y_{d}(t)=\sqrt{r_{g}^{\prime }(t)}(h(t))^{2a/c}
(U_{u})_{L}^{\dagger}Z(t)(U_{u})_{L}y_{d}(t_{0}),\label{ant1}
\end{eqnarray}
where the $(a,b,c)$ are equal to $(0,2,2/3)$, $(0,1,1/3)$,
$(1,1,-1)$ in the MSSM, DHS and SM, respectively,
\begin{eqnarray}
h(t) &=&\exp (\frac{1}{(4\pi)^{2}}\frac{3c}{2}
\int_{t_{0}}^{t}m_{t}^{2}(\tau)d\tau)  \nonumber \\
&=&\left(\frac{1}{1-\frac{3(b+2)}{(4\pi)^{2}}
m_{t}^{2}(t_{0})
\int_{t_{0}}^{t}r_{g}(\tau)d\tau}\right)^{\frac{c}{2(b+2)}}.
\label{htt}
\end{eqnarray}
and
\[
\left[Z(t)\right] _{ij}=\delta _{ij}+(h(t)-1)\delta _{i3}\delta _{j3}.
\]
These solutions Eqs.~(\ref{ant1}) do depend on the energy through the
overall factors $r_{g}(t),$ $r_{g}^{\prime }(t)$ (see Appendix)
and the matrix $Z(t)$.

The procedure Ref.~\cite{ref5}, to obtain the energy dependence of the
masses for the down sector is the following:\\[4pt]
Step 1.- Differentiate with respect to $t$ the following equation
\[
(U_{u})_{L}y_{d}(t)y_{d}(t)^{\dagger }(U_{u})_{L}^{\dagger }
=r_{g}^{^{\prime}}(t)(h(t))^{(2\alpha _{3}^{d}/\alpha
_{2}^{d})}Z(t)(U_{u})_{L}y_{d}(t_{0})y_{d}(t_{0})^{\dagger
}(U_{u})_{L}^{\dagger }Z(t),
\]
which can be written in this way
\[
(U_{u})_{L}y_{d}(t)y_{d}(t)^{\dagger }
(U_{u})_{L}^{\dagger}=V_{\mbox{\scriptsize CKM}}(t)M_{d}^{2}(t)
V_{\mbox{\scriptsize CKM}}^{\dagger }(t)
\]
where $M_{d}^{2}(t)$ is the diagonal matrix with the squares of the
physical down quarks Yukawa couplings on the diagonal, that become the
squares of the down quark masses after the spontaneous symmetry
breaking. Next we obtain the following matrix differential equation
\begin{eqnarray}
&&V_{\mbox{\scriptsize CKM}}^{\dagger }(t)
\frac{dV_{\mbox{\scriptsize CKM}}}{dt}=
(M_{d}^{2})^{-1}V_{\mbox{\scriptsize CKM}}^{\dagger
}(t)\frac{{dV_{\mbox{\scriptsize CKM}}}}{dt}M_{d}^{2}  \nonumber \\
&&-(M_{d}^{2})^{-1}V_{\mbox{\scriptsize CKM}}^{\dagger }(t)\frac{{d}}{dt}
((U_{u})_{L}y_{d}(t)y_{d}(t)^{\dagger }(U_{u})_{L}^{\dagger
})V_{\mbox{\scriptsize CKM}}(t)+(M_{d}^{2})^{-1}\frac{dM_{d}^{2}}{dt}.
\label{mde}
\end{eqnarray}
Eq.~(\ref{mde}) becomes simpler after using the following relation (see
Eq.~(\ref{ant1}))
\begin{eqnarray*}
&&(M_{d}^{2})^{-1}V_{\mbox{\scriptsize CKM}}^{\dagger }(t)\frac{{d}}{dt}
((U_{u})_{L}y_{d}(t)y_{d}(t)^{\dagger }(U_{u})_{L}^{\dagger })
V_{\mbox{\scriptsize CKM}}(t) \\
&=&\frac{h^{^{\prime }}}{h}({\bf R}^{\dagger }{\bf R}+(M_{d}^{2})^{-1}{\bf R}
^{\dagger }{\bf R}M_{d}^{2})+\frac{d\ln (r_{g}^{^{\prime
}}(t)(h(t))^{(2\alpha _{3}^{d}/\alpha _{2}^{d})})}{dt}I
\end{eqnarray*}
where the vector ${\bf R}=(V_{td},V_{ts},V_{tb}).$\\[3pt]
Step 2.- Extract the differential equations for the diagonal matrix
elements of Eq.~(\ref{mde}):
\begin{eqnarray}
\frac{d}{dt}\ln m_{d}(t) &=&\frac{1}{2}\frac{d\ln \left[
r_{g}^{^{\prime }}(t)(h(t))^{4a/c}\right] }{dt}+\left[\frac{d}{dt}\ln
h(t) \right] \left| V_{td}(t) \right| ^{2},
\nonumber \\
\frac{d}{dt}\ln m_{s}(t) &=&\frac{1}{2}\frac{d\ln \left[
r_{g}^{^{\prime }}(t)(h(t))^{4a/c}\right] }{dt}+\left[\frac{d}{dt}\ln
h(t) \right] \left| V_{ts}(t) \right| ^{2},
\nonumber \\
\frac{d}{dt}\ln m_{b}(t) &=&\frac{1}{2}\frac{d\ln \left[
r_{g}^{^{\prime }}(t)(h(t))^{4a/c}\right] }{dt}+\left[\frac{d}{dt}\ln
h(t) \right] \left| V_{tb}(t) \right| ^{2}.
\label{mb}
\end{eqnarray}
Step 3.- From the off diagonal matrix elements of Eq.~(\ref{mde}) we
deduce equations for the squares of the CKM matrix elements
$|V_{ij}|^2$. We solve equations for the $|V_{td}|^2$, $|V_{ts}|^2$
and $|V_{tb}|^2$ matrix elements that are needed in Eq.~(\ref{mb}):
\begin{eqnarray}
\frac{d}{dt}\left| V_{td}(t) \right| ^{2} &=&-2\left[\frac{d}{dt
}\ln h(t) \right] \left| V_{td}(t) \right|
^{2}(1-|V_{td}(t) |^{2}),  \nonumber \\
\frac{d}{dt}|V_{tb}(t) |^{2} &=&2\left[\frac{d}{dt}\ln h(t)
 \right] |V_{tb}(t) |^{2}(1-|V_{tb}(t)|^{2}),  \nonumber \\
\frac{d}{dt}\left| V_{ts}(t) \right| ^{2} &=&-2\left[\frac{d}{dt
}\ln h(t) \right] \left| V_{ts}(t) \right|
^{2}\left[\left| V_{tb}(t) \right| ^{2}-\left| V_{td}(t)
 \right| ^{2}\right].
\end{eqnarray}
Obtaining
\begin{eqnarray}
|V_{td}(t)|^{2} &=&\frac{|V_{td}^{0}|^{2}}{h^{2}(t)
+(1-h^{2}(t))|V_{td}^{0}|^{2}}
,\,\,\,\,\,\,\,\,\,\,\,\,\,\,|V_{tb}\left(t\right) |^{2}=\frac{
|V_{tb}^{0}|^{2}h^{2}(t)}{1+|V_{tb}^{0}|^{2}(h^{2}(t) -1)},\nonumber \\
|V_{ts}(t)|^{2} &=&\frac{|V_{ts}^{0}|^{2}}{|V_{tb}^{0}|^{2}
\left[1-|V_{td}^{0}|^{2}\right] }\frac{h^{2}(
t) }{\left[h^{2}(t) +\frac{1-|V_{tb}^{0}|^{2}}{
|V_{tb}^{0}|^{2}}\right] \left[h^{2}(t) +\frac{|V_{td}^{0}|^{2}}
{\left[1-|V_{td}^{0}|^{2}\right] }\right] },  \label{solv}
\end{eqnarray}
where
\[
\,\,\,V_{ij}^{0}=V_{ij}(t_{0}).
\]
Step 4.- Finally, Eqs.~(\ref{solv}) are used to solve in an analytical way
the differential equations~(\ref{mb}).

The formulas which give the explicit energy dependence of the quark masses
for the down sector are
\begin{eqnarray}
\frac{m_{d}(t) }{m_{d}(t_{0}) } &=&\sqrt{
r_{g}^{^{\prime }}(t)}\frac{(h(t))^{(c+2a)/c}}{\sqrt{h^2(t)
+|V_{td}^{0}|^{2}(1-h^2(t)) }},
\nonumber \\
\frac{m_{s}(t)}{m_{s}(t_{0})} &=&\sqrt{
r_{g}^{^{\prime }}(t)}\frac{\sqrt{h^2(t)
+|V_{td}^{0}|^{2}(1-h^2(t))}}{\sqrt{
1+|V_{tb}^{0}|^{2}(h^2(t)-1) }}(h(t))^{2a/c},  \nonumber \\
\frac{m_{b}(t) }{m_{b}(t_{0}) } &=&\sqrt{
r_{g}^{^{\prime }}(t)}\sqrt{1+|V_{tb}^{0}|^{2}(h^2(t)-1) \,}(h(t))^{2a/c}.
\end{eqnarray}
\begin{figure}[h]
\centerline{\epsfig{file=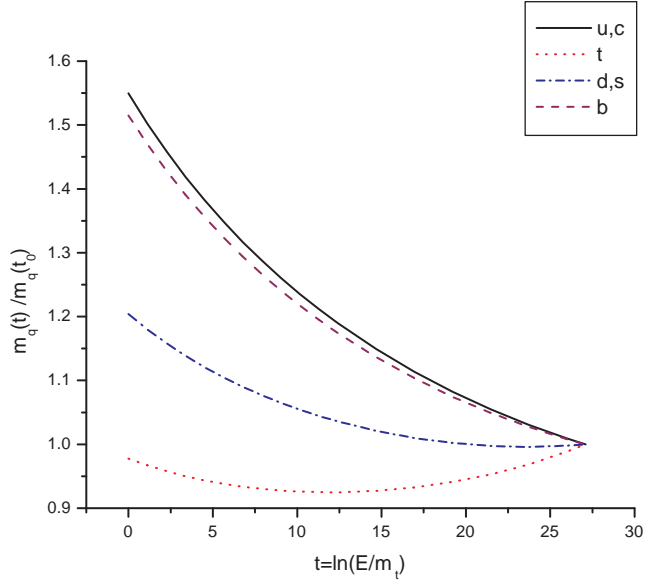}}
\vspace{10pt}
\caption{Evolution of the quark masses in the Standard Model}
\label{fig1}
\end{figure}
The results presented in this paper together with those of
Ref.~\cite{ref0} form the complete set of the renormalization group
evolution predictions for the observables derived from the quark
Yukawa couplings. The consistent approximation scheme based on the
hierarchy of the quark masses and the CKM matrix is strictly observed
in all the derivations and it is shown that the final results for the
ratios of the quark masses and the CKM matrix depend only on one
function of energy in agreement with the theorem presented in
Ref.~\cite{ref0}.

As an illustration we show in Fig.~1 the evolution of the quark masses
for the energy range from $E=m_t$ to $E=10^{14}$~GeV for the Standard
Model.

The explicit form of the evolution given here can be very useful for
the phenomenological analysis of the models that specify the
properties of the Yukawa interactions at the GU scale.

We acknowledge the financial support from CONACYT (M\'{e}xico)
projects 3512P-E9608 and 26247E. S.R.J.W.  also thanks to
``Comisi\'{o}n de Operaci\'{o}n y Fomento de Actividades
Acad\'{e}micas'' (COFAA) from Instituto Polit\'{e}cnico Nacional.

\section*{Appendix}
\renewcommand{\theequation}{A\arabic{equation}}
\setcounter{equation}{0}
\begin{eqnarray}
\beta _{l}^{(1)} &=&\alpha _{1}^{l}(t)+\alpha _{2}^{l}H_{u}^{(1)}
+\alpha_{3}^{l}\mbox{Tr}(H_{u}^{(1)})+\alpha _{4}^{l}H_{d}^{(1)}
+\alpha _{5}^{l}\mbox{Tr}(H_{d}^{(1)})  \nonumber \\
&+&\alpha _{6}^{l}H_{e}^{(1)}+\alpha _{7}^{l}\mbox{Tr}(H_{e}^{(1)})
+\alpha_{8}^{l}H_{\nu }^{(1)}+\alpha _{9}^{l}
\mbox{Tr}(H_{\nu}^{(1)}),\,\,\,\,H_{l}^{(1)}=y_{l}y_{l}^{\dagger }.
\end{eqnarray}
$l=u,d,e,\nu.$ The $\alpha _{1}^{l}(t)$ and the $\alpha _{i}^{l},$
$i=2,...,9$ are given in Tables 2 and 3
\begin{eqnarray}
\displaystyle r_{g}(t) &=&\exp (\frac{2}{(4\pi)^{2}}
\int_{t_{0}}^{t}\alpha _{1}^{u}(\tau)d\tau) =\Pi _{k=1}^{k=3}
\left[\frac{g_{k}^{2}(t_{i}) }{g_{k}^{2}(t) }\right]^{
\frac{c_{k}}{b_{k}}},  \nonumber \\
\displaystyle r_{g}^{\prime }(t) &=&\exp (\frac{2}{(4\pi
)^{2}}\int_{t_{0}}^{t}\alpha _{1}^{d}(\tau)d\tau) =\Pi
_{k=1}^{k=3}\left[\frac{g_{k}^{2}(t_{i}) }{g_{k}^{2}(
t) }\right] ^{\frac{c_{k}^{\prime }}{b_{k}}},
\end{eqnarray}
where
\[
g_{k}(t)=\frac{g_{k}(t_{0})}{\sqrt{1-\frac{
2b_{k}g_{k}^{2}(t_{0})(t-t_{0})}{(4\pi)^{2}}}}.
\]
The coefficients $b_k$, $c_k$ and $c^{\prime}_k$ are defined in Table~1.
\begin{table}[h]
\caption{The parameters for the various models.}
\label{table1}
\begin{tabular}{|c|c|c|c|c|c|c|c|c|c|}
\hline
Model & $b_{1}$ & $b_{2}$ & $b_{3}$ & $c_{1}$ & $c_{2}$ & $c_{3}$ & $
c_{1}^{\prime }$ & $c_{2}^{\prime }$ & $c_{3}^{\prime }$ \\ \hline
MSSM & $\vphantom{\frac{\frac{A}{B}}{\frac{A}{B}}
\frac{\frac{A}{B}}{\frac{A}{B}}}
\frac{33}{5}$ & $1$ & $-3$ & $\frac{13}{15}$ & $3$ & $\frac{16}{3}$
& $\frac{7}{15}$ & $3$ & $\frac{16}{3}$ \\ \hline
DHM & $\vphantom{\frac{\frac{A}{B}}{\frac{A}{B}}
\frac{\frac{A}{B}}{\frac{A}{B}}}
\frac{21}{5}$ & $-3$ & $-7$ & $\frac{17}{20}$ & $\frac{9}{4}$ & $8$ &
$\frac{1}{4}$ & $\frac{9}{4}$ & $8$ \\ \hline
SM & $\vphantom{\frac{\frac{A}{B}}{\frac{A}{B}}
\frac{\frac{A}{B}}{\frac{A}{B}}}
\frac{41}{10}$ & $\frac{-19}{6}$ & $-7$ & $\frac{17}{20}$ & $\frac{9}{4
}$ & $8$ & $\frac{1}{4}$ & $\frac{9}{4}$ & $8$ \\ \hline
\end{tabular}
\end{table}
\begin{table}[h]
\caption{The coefficients $\alpha_1^l$ for various models.}
\label{table2}
\begin{tabular}{|l|l|l|}
\hline
& $\text{SM and DHM}$ & \thinspace \thinspace \thinspace \thinspace
\thinspace \thinspace \thinspace \thinspace \thinspace \thinspace \thinspace
\thinspace \thinspace \thinspace \thinspace MSSM \\ \hline
$\vphantom{\frac{\frac{A}{B}}{\frac{A}{B}}
\frac{\frac{A}{B}}{\frac{A}{B}}}
\alpha _{1}^{u}(t)=$ & $-(\frac{17}{20}g_{1}^{2}+\frac{9}{4}
g_{2}^{2}+8g_{3}^{2})$ & $-(\frac{13}{15}g_{1}^{2}+3g_{2}^{2}+\frac{16}{3}
g_{3}^{2})$ \\ \hline
$\vphantom{\frac{\frac{A}{B}}{\frac{A}{B}}
\frac{\frac{A}{B}}{\frac{A}{B}}}
\alpha _{1}^{d}(t)=$ & $-(\frac{1}{4}g_{1}^{2}+\frac{9}{4}
g_{2}^{2}+8g_{3}^{2})$ & $-(\frac{7}{15}g_{1}^{2}+3g_{2}^{2}+\frac{16}{3}
g_{3}^{2})$ \\ \hline
$\vphantom{\frac{\frac{A}{B}}{\frac{A}{B}}
\frac{\frac{A}{B}}{\frac{A}{B}}}
\alpha _{1}^{e}(t)=$ & $-(\frac{9}{20}g_{1}^{2}+\frac{9}{4}g_{2}^{2})$ & $-(
\frac{9}{5}g_{1}^{2}+3g_{2}^{2})$ \\ \hline
$\vphantom{\frac{\frac{A}{B}}{\frac{A}{B}}
\frac{\frac{A}{B}}{\frac{A}{B}}}
\alpha _{1}^{\nu }(t)=$ & $-(\frac{9}{20}g_{1}^{2}+\frac{9}{4}g_{2}^{2})$ &
$-(\frac{3}{5}g_{1}^{2}+3g_{2}^{2})$ \\ \hline
\end{tabular}
\end{table}
\begin{table}[h]
\caption{The coefficients $\alpha_k^l$ for various models;
the constants $(a,b,c)$ are equal to $(0,2,2/3)$, $(0,1,1/3)$,
$(1,1,-1)$ in the MSSM, DHS and SM.}
\label{table3}
\begin{tabular}{|c|c|c|c|c|c|c|c|c|}
\hline
$l\vphantom{\frac{\frac{A}{B}}{\frac{A}{B}}
\frac{\frac{A}{B}}{\frac{A}{B}}}
$ & $\alpha _{2}^{l}$ & $\alpha _{3}^{l}$ & $\alpha _{4}^{l}$ & $\alpha
_{5}^{l}$ & $\alpha _{6}^{l}$ & $\alpha _{7}^{l}$ & $\alpha _{8}^{l}$ & $
\alpha _{9}^{l}$ \\ \hline
$u\vphantom{\frac{\frac{A}{B}}{\frac{A}{B}}
\frac{\frac{A}{B}}{\frac{A}{B}}}
$ & $\frac{3}{2}b$ & $3$ & $\frac{3}{2}c$ & $3a$ & $0$ & $a$ & $0$ & $1$
\\ \hline
$d\vphantom{\frac{\frac{A}{B}}{\frac{A}{B}}
\frac{\frac{A}{B}}{\frac{A}{B}}}
$ & $\frac{3}{2}c$ & $3a$ & $\frac{3}{2}b$ & $3$ & $0$ & $1$ & $0$ & $a$
\\ \hline
$e\vphantom{\frac{\frac{A}{B}}{\frac{A}{B}}
\frac{\frac{A}{B}}{\frac{A}{B}}}
$ & $0$ & $3a$ & $0$ & $3$ & $\frac{3}{2}b$ & $1$ & $\frac{3}{2}c$ & $a$
\\ \hline
$\nu\vphantom{\frac{\frac{A}{B}}{\frac{A}{B}}
\frac{\frac{A}{B}}{\frac{A}{B}}}
 $ & $0$ & $3$ & $0$ & $3a$ & $\frac{3}{2}c$ & $a$ & $\frac{3}{2}b$ & $1$
\\ \hline
\end{tabular}
\end{table}

\end{document}